\documentclass[
 reprint,
superscriptaddress,
 amsmath,amssymb,
 aps,
]{revtex4-2}
\usepackage{amsmath, bm}
\usepackage{epsfig}
\usepackage{color}
\usepackage{soul}
\usepackage{lineno,hyperref}
\usepackage{multirow}
\newcommand{\sNN}{$\sqrt{s_{_\mathrm{NN}}}$ }

\usepackage{tabularx}
\usepackage{graphicx}
\usepackage{dcolumn}
\usepackage{bm}

\usepackage{slashed}
\usepackage{comment}

\begin{document}

\preprint{APS/123-QED}

\title{Production of muonic kaon atoms at high-energy colliders}

\author{Xiaofeng Wang}
\email{xiaofeng\_wang@ustc.edu.cn}
\affiliation{Department of Modern Physics, University of Science and Technology of China, Hefei, Anhui
230026, China}

\author{Zebo Tang}
\email{zbtang@ustc.edu.cn}
\affiliation{Department of Modern Physics, University of Science and Technology of China, Hefei, Anhui 
230026, China}

\author{Zhangbu Xu}
\email{zxu22@kent.edu}
\affiliation{Physics Department, Kent State University, Kent, OH 44242, USA}
\affiliation{Physics Department, Brookhaven National Laboratory, Upton, NY11973, USA}

\author{Chi Yang}
\email{chiyang@sdu.edu.cn}\affiliation{Key Laboratory of Particle Physics and Particle Irradiation (MOE), Institute of Frontier and Interdisciplinary Science, Shandong University, Qingdao, China 266237}

\author{Wangmei Zha}
\email{first@ustc.edu.cn}
\affiliation{Department of Modern Physics, University of Science and Technology of China, Hefei, Anhui 
230026, China}

\author{Yifei Zhang}
\email{ephy@ustc.edu.cn}
\affiliation{Department of Modern Physics, University of Science and Technology of China, Hefei, Anhui 
230026, China}

\begin{abstract}
We develop a framework for the formation of exotic muonic kaon atoms ($K\mu$) in semileptonic $D^{0}$ decays, using the effective weak Hamiltonian, a helicity-based treatment of the leptonic current, and a nonrelativistic bound-state projection. The resulting branching ratio, $\mathrm{BR}(D^{0}\!\to(K\mu )\nu_{\mu})=2.29\times10^{-10}$, is implemented in a ROOT-based code to estimate yields at RHIC, LHC, and STCF. We show quantitatively that $K\mu$ atoms—also produced through coalescence in the quark--gluon plasma (QGP)—provide a sensitive probe of low-momentum primordial muons and early-time electromagnetic radiation, offering complementary constraints in an otherwise unexplored phase space for thermal dilepton and photon emission. Newly estimated dissociation cross sections in detector material indicate that secondary‑vertex reconstruction should be experimentally  feasible, allowing clean experimental identification of the atoms. Projected yields from QGP coalescence in LHC and RHIC heavy-ion collisions, and from $D^{0}$ decays in LHC high-luminosity $p+p$ collisions indicate that the first observation of $K\mu$ atoms is within reach.
\end{abstract}

\maketitle


\section{Introduction}

Muonic atoms are exotic Coulomb bound states in which an electron in an ordinary atom is replaced by a muon. Owing to the muon mass, $m_\mu \simeq 207\,m_e$, the characteristic Bohr radius is reduced by the same factor, bringing the lepton wave function deep inside the hadronic/nuclear charge distribution~\cite{CREMA:2025zpo,PhysRevLett.51.1633}. As a consequence, atomic energy levels, hyperfine splittings, and transition rates in muonic atoms become unusually sensitive to finite-size and internal-structure effects of the bound hadron or nucleus, providing clean and quantitative access to electromagnetic form factors~\cite{Persson_2025_BW}, charge radii~\cite{Antognini_2021_SciPostProc}, and polarizabilities~\cite{Gorchtein_2025_arXiv,LiMuli_2022_JPhysG}. Historically, precision spectroscopy of muonic systems has enabled some of the most stringent determinations of nuclear charge radii~\cite{Antognini_2021_SciPostProc} and has stimulated broad interest in testing bound-state QED in strong fields~\cite{Okumura_2023_PRL130,cheng1981pimu,Baym:1993ae} and in disentangling hadronic-structure corrections~\cite{Eskin_2017_EPJWC}. From this perspective, extending muonic-atom studies from the well-established muonic hydrogen/deuterium and muonic helium to muon--meson atoms opens a qualitatively new window: mesons are composite QCD bound states with short lifetimes and distinct internal dynamics, so producing and identifying such ``muonic mesonic atoms'' would provide a rare laboratory for probing meson structure in an atomic setting.

Despite this motivation, muonic meson atoms remain largely unexplored experimentally. To date, only one muonic meson atom, the $(\pi\mu)_{\rm atom}$, has been experimentally observed in kaon decay experiments~\cite{Coombes:1976hi,Aronson:1982bz}. In those measurements, $(\pi\mu)_{\rm atom}$ was formed in semileptonic $K_L^0$ decays, $K_L^0\to(\pi\mu)_{\rm atom}\nu$, where the pion and muon can be produced with ultra-small relative momentum and bind via the Coulomb interaction. A long evacuated decay region was essential to suppress secondary interactions and to preserve such a fragile bound state, while a thin downstream foil was used to dissociate the electrically neutral atom into a detectable $\pi$--$\mu$ ``atomic pair''. After breakup, the two tracks emerge nearly collinear with (approximately) the same velocity, yielding a distinctive momentum correlation, and muon identification further suppresses pion/electron contamination. This experiment  already illustrates that mesonic muonic atoms are rare and experimentally delicate objects.

For the kaon--muon system $(K\mu)_{\rm atom}$, one might consider a beamline-style search based on a heavy-flavor parent, e.g.\ $D^{0}\to (K\mu)_{\rm atom}\nu_{\mu}$, in analogy with breakup-based identification strategies. In practice, however, such an approach is not viable because the $D^0$ has a very short proper decay length, $c\tau\simeq 123~\mu\mathrm{m}$~\cite{ParticleDataGroup:2024cfk}, corresponding to sub-mm to mm decay lengths for typical collider boosts, far too short for a beamline-style transported decay region and a controlled downstream breakup stage as in the $K_L\rightarrow(\pi\mu)\nu$ process. Consequently, realistic searches for $(K\mu)_{\rm atom}$ must be carried out directly in collider events. This naturally points to two complementary directions: coalescence formation driven by Coulomb final-state interactions in high-energy particle collisions, and formation in heavy-flavor decays, where the decay channel itself provides a well-defined source of correlated $K$--$\mu$ pairs.

In this spirit, one well-studied possibility is coalescence formation driven by Coulomb final-state interactions in relativistic heavy-ion collisions (HIC)~\cite{Baym:1993ae}. The large particle multiplicities and the late-stage evolution of the system enhance the probability that oppositely charged constituents with sufficiently small relative momentum form a bound state. In our recent work~\cite{Wang:2024njn}, we investigated the formation of muonic atoms in HIC and showed that incorporating Coulomb correlations from freeze-out to atom formation can significantly impact the expected yields. At the same time, $(K\mu)_{\rm atom}$ measurements can be used in the reverse direction: because atomic formation requires a nearly comoving kaon--muon pair, the $(K\mu)_{\rm atom}$ yield is tightly correlated with the abundance of primordial low-$p_T$ muons from thermal QGP radiation~\cite{Baym:1993ae,Kapusta:1998fh,Wang:2024njn}. Therefore, observing $(K\mu)_{\rm atom}$ provides an experimentally accessible way to constrain the thermal-muon yield, which in turn helps constrain the QGP electromagnetic emissivity and the medium's temperature and space--time evolution~\cite{STAR:2016use,PHENIX:2022rsx,ALICE:2023jef,STAR:2024bpc}.

To follow up on the two complementary directions outlined above, coalescence describes $(K\mu)_{\rm atom}$ formation in the bulk collision environment, while a second, channel-defined source can arise from heavy-flavor semileptonic decays. To our knowledge, however, a dedicated quantitative estimate of $(K\mu)_{\rm atom}$ formation in $D^{0}\to (K\mu)_{\rm atom}\nu_{\mu}$ has been lacking. This paper provides such a complementary proposal by developing a quantitative description of $(K\mu)_{\rm atom}$ formation in the clean semileptonic channel $D^{0}\to (K\mu)_{\rm atom}\nu_{\mu}$, and evaluating its contribution alongside the HIC coalescence mechanism. Yield projections are presented for RHIC, the LHC, and STCF. Crucially, newly estimated dissociation of $(K\mu)_{\rm atom}$ in detector material is incorporated to translate formation rates into measurable surviving (or breakup) signals and to motivate practical search strategies based on the characteristic near-comoving ``atomic-pair'' kinematics after dissociation.

The remainder of this paper is organized as follows. 
Section II presents the theoretical framework for $D^0\to (K\mu)_{\rm atom}\nu_\mu$, including the free semileptonic decay amplitude, the bound-state projection, production-yield estimates, and atomic dissociation in detector material. 
Section III presents the predicted branching ratios and yields, compares the decay and coalescence production mechanisms, and discusses the implications for thermal-muon sensitivity and experimental feasibility. 
Section IV summarizes the main conclusions.

\section{Theoretical Framework and Methodology}

In this section, we present the theoretical framework for the decay-driven production of muonic kaon atoms in the semileptonic channels
\begin{align}
D^0 &\to K^- \mu^+ \nu_\mu \,, \\
D^0 &\to (K\mu)_{\rm atom}\,\nu_\mu \,,
\end{align}
where $(K\mu)_{\rm atom}$ denotes the Coulomb bound state of $\mu^+$ and $K^-$. 
Our strategy is to first construct the free three-body decay amplitude using the effective weak Hamiltonian, the hadronic form factors, and the helicity-basis leptonic current, and then project the near-threshold $K\mu$ pair onto a nonrelativistic Coulomb bound state. 
This framework allows us to evaluate the atomic branching ratio and, together with facility-specific charm yields and detector material budgets, to estimate the corresponding production rates and experimental signatures at RHIC, the LHC, and STCF.

Throughout this work, we use natural units $\hbar = c = 1$. The numerical values of the particle masses, the Fermi constant $G_F$, and the CKM matrix element $V_{cs}$ are taken from the Particle Data Group~\cite{ParticleDataGroup:2024cfk}.

\subsection{Free semileptonic decay amplitude for $D^0\to K^-\mu^+\nu_\mu$}

At the hadronic scale, the semileptonic decay
\[
D^0(c\bar u) \to K^-(\bar us )\, \mu^+ \nu_\mu
\]
is described by the effective four-fermion interaction
\begin{equation}
\mathcal{H}_{\text{eff}}
= \frac{G_F}{\sqrt{2}}\, V_{cs}\,
   \big[\, \bar s\, \gamma^\mu (1 - \gamma_5)\, c \,\big]\,
   \big[\, \bar \mu\, \gamma_\mu (1 - \gamma_5)\, \nu_\mu \,\big] + \text{h.c.}
\end{equation}
The decay amplitude is then written as
\begin{equation}
\begin{aligned}
\mathcal{M}
= \frac{G_F}{\sqrt{2}}\, V_{cs}\; H^\mu\, L_\mu \,,
\end{aligned}
\end{equation}
where $G_F$ is the Fermi coupling constant, $V_{cs}$ is the relevant Cabibbo-Kobayashi-Maskawa (CKM) matrix element, and $H^\mu$ and $L_\mu$ represent the hadronic and leptonic currents, respectively, given by:
\begin{equation}
\begin{aligned}
H^\mu = \langle K^-|\bar s\, \gamma^\mu (1-\gamma_5)\, c|D^0\rangle,\\
L_\mu = \bar u_\nu(p_\nu)\,\gamma_\mu(1-\gamma_5)\,v_\mu(p_\mu) \,.
\end{aligned}
\end{equation}

Since both $D^0$ and $K^-$ are pseudoscalar mesons, only the vector part contributes to the hadronic matrix element. It can be parameterized in terms of two form factors, $f_+(q^2)$ and $f_0(q^2)$:
\begin{equation}
\begin{aligned}
H^\mu \equiv \langle K^-(p_K)|\bar s\,\gamma^\mu\, c|D^0(p_D)\rangle\\
=f_+(q^2)\left[(p_D+p_K)^\mu - \frac{m_D^2-m_K^2}{q^2}\,q^\mu\right]\\
+ f_0(q^2)\,\frac{m_D^2-m_K^2}{q^2}\,q^\mu,
\label{eq:Hmu_fplus_f0}
\end{aligned}
\end{equation}
where
\[
q^\mu = p_D^\mu - p_K^\mu,
\qquad
q^2 = (p_D - p_K)^2.
\]

The leptonic current is given by
\begin{equation}
L^\mu = \bar u_\nu(p_\nu)\, \gamma^\mu (1-\gamma^5) v_\mu(p_\mu),
\end{equation}
and the corresponding spin-summed leptonic tensor can be written as
\begin{equation}
L^{\mu\nu} = \sum_{\text{spins}} L^\mu\, (L^\nu)^\ast
= \sum_{h_\mu=\pm1} L^\mu(h_\mu)\, L^{\nu}(h_\mu)^\ast.
\end{equation}

In the $D^0$ rest frame, $p_D^\mu = (m_D,\bm{0})$, the differential decay width takes the form
\begin{equation}
\frac{d^2\Gamma}{dq^2\,d\cos\theta_\ell}
=
\frac{G_F^2 |V_{cs}|^2}{2}\;
\frac{p_K\,p_\mu}{64\pi^3\,m_D^2\,\sqrt{q^2}}\;
\frac{1}{2m_D}\;
M_{\text{free}}^{2},
\end{equation}
where $M_{\text{free}}^{2} \equiv H_\mu (H_\nu)^\ast L^{\mu\nu}$ consistently denotes the spin-summed squared matrix element for the free decay throughout this work.

The total free width is obtained by integrating over $q^2$ and $\cos\theta_\ell$:
\begin{equation}
\Gamma(D^0 \to K^- \mu^+ \nu_\mu)
=
\int_{q^2_{\min}}^{q^2_{\max}} dq^2
\int_{-1}^{+1} d\cos\theta_\ell\;
\frac{d^2\Gamma}{dq^2 d\cos\theta_\ell}.
\end{equation}
The corresponding branching ratio is
\begin{equation}
\text{BR}(D^0 \to K^- \mu^+ \nu_\mu)
=
\Gamma(D^0 \to K^- \mu^+ \nu_\mu)\,
\frac{\tau_{D^0}}{\hbar},
\end{equation}
yielding an estimated value of $\text{BR} \approx 0.034$ based on current lattice and experimental inputs~\cite{ParticleDataGroup:2024cfk,BESIII:2024slx,Chakraborty:2021qav}.

\subsection{Bound-state projection for $D^0\to (K\mu)_{\rm atom}\nu_\mu$}

To describe the formation of the atomic bound state in the decay $D^0 \to (K\mu)_{\text{atom}} \nu_\mu$, we develop a projection framework based on the separation of physical scales. A fundamental observation is that the $D^0$ meson has a proper lifetime of $\tau \approx 410.1$~fs, which is approximately eleven orders of magnitude longer than the expansion time of the transient hadronic medium ($\sim 10$~fm/$c$). This ensures that the $D^0$ meson escapes the collision environment and decays in vacuum. Thus, the $(K\mu)_{\text{atom}}$ formation is a vacuum process governed by the total atomic mass $M_A \approx m_K + m_\mu$. In the $D^0$ rest frame, the center-of-mass momentum $|\bm{p}_A|$ and energy $E_A$ of the outgoing atom are:
\begin{equation}
|\bm{p}_A| = \frac{m_D^2 - M_A^2}{2m_D}, \quad E_A = \frac{m_D^2 + M_A^2}{2m_D}.
\end{equation}

In general, an atomic bound state $| A_{nl}(\bm{P}) \rangle$ with total momentum $\bm{P}$, principal quantum number $n$, and orbital angular momentum $l$ is represented in second quantization as:
\begin{equation}
\begin{aligned}
| A_{nl}(\bm{P}) \rangle = &\sqrt{2M_A} \int \frac{d^3\bm{k}}{(2\pi)^3} \frac{\tilde{\psi}_{nl}(\bm{k})}{\sqrt{2E_\mu 2E_K}} \\
&\times | \mu^+(\bm{p}_\mu), K^-(\bm{p}_K) \rangle,
\end{aligned}
\end{equation}
where $\bm{k}$ is the relative momentum between the constituents, $\tilde{\psi}_{nl}(\bm{k})$ is the momentum-space wavefunction, and $\bm{p}_\mu, \bm{p}_K$ are the laboratory momenta satisfying $\bm{p}_\mu + \bm{p}_K = \bm{P}$. The transition matrix element $\mathcal{M}_{\text{bound}}$ for the atomic channel is given by:
\begin{equation}
\begin{aligned}
\mathcal{M}_{\text{bound}} &= \langle A_{nl}(\bm{P}), \nu_\mu(p_\nu) | \mathcal{H}_{\text{eff}} | D^0(p_D) \rangle \\
&\simeq \sqrt{2M_A} \int \frac{d^3\bm{k}}{(2\pi)^3} \frac{\tilde{\psi}_{nl}^*(\bm{k})}{\sqrt{2E_\mu 2E_K}} \mathcal{M}_{\text{free}}(\bm{k}).
\end{aligned}
\end{equation}

In the non-relativistic limit, the wavefunction $\tilde{\psi}_{nl}(\bm{k})$ is sharply peaked at $|\bm{k}| \simeq 0$, with a characteristic width $\sim \alpha \mu_{\text{red}}$ much smaller than the constituent masses. In this threshold region, the energies of the kaon and muon satisfy $E_{K,\mu} = \sqrt{m_{K,\mu}^2 + \bm{k}^2} \approx m_{K,\mu}$. This justifies replacing the dynamical energy factors in the denominator with static masses. Furthermore, since the free decay amplitude $\mathcal{M}_{\text{free}}(\bm{k})$ varies smoothly, it can be evaluated at the threshold ($s_{\mu K} = M_A^2$) and factored out of the integral:
\begin{equation}
\begin{aligned}
\mathcal{M}_{\text{bound}} &\simeq \sqrt{2M_A} \frac{\mathcal{M}_{\text{free}}|_{s_{\mu K}=M_A^2}}{\sqrt{2m_\mu 2m_K}} \int \frac{d^3\bm{k}}{(2\pi)^3} \tilde{\psi}_{nl}^*(\bm{k}) \\
&= \sqrt{2M_A}\,\psi_{nl}(0)\, \frac{\mathcal{M}_{\text{free}}|_{s_{\mu K}=M_A^2}}{\sqrt{2m_\mu\,2m_K}},
\end{aligned}
\end{equation}
where $\psi_{nl}(0)$ is the real-space wavefunction at the origin. Since $\psi_{nl}(0)$ vanishes for all states with $l > 0$ due to the centrifugal barrier, only the $nS$ states ($l=0$) contribute to the formation.

In the numerical implementation, the spin-summed contraction $H_\alpha H_\beta^\ast L^{\alpha\beta}$ is written as:
\begin{equation}
\begin{aligned}
H_\alpha H_\beta^\ast L^{\alpha\beta} = &\, 4\,\frac{m_\mu}{M_A}\,\bigl(m_D^{2}-M_A^{2}\bigr) \\
&\times \Bigl[\,2m_K + m_\mu + g(q^{2})\,m_\mu\,\Bigr]^{2} \lvert f_{+}(q^{2})\rvert^{2},
\label{eq:HHLL_D0_muK_code}
\end{aligned}
\end{equation}
where $g(q^{2})$ is the $q^{2}$-dependent correction from the full helicity structure. The $1S$ wavefunction at the origin is determined by the reduced mass $\mu_{\text{red}} = (m_\mu m_K)/(m_\mu + m_K)$ and the fine-structure constant $\alpha$ as $|\psi_{1S}(0)|^2 = (\mu_{\text{red}} \alpha)^3 / \pi$. The final decay width for the $1S$ state is:
\begin{equation}
\begin{aligned}
\Gamma&(D^0 \to (K\mu)_{1S}\,\nu_\mu) = \frac{|\bm{p}_A|\,M_A}{16\pi\,m_D^2\,m_\mu\,m_K} \\
&\times |\psi_{1S}(0)|^2 \frac{G_F^2 |V_{cs}|^2}{2} M^2_{\text{free}}(M_A).
\label{eq:Gamma_atom_final}
\end{aligned}
\end{equation}

For the full tower of $nS$ states, the sum over all $n$ gives:
\begin{equation}
\sum_{n=1}^{\infty} |\psi_{nS}(0)|^2 = \frac{(\mu_{\text{red}}\alpha)^3}{\pi} \sum_{n=1}^{\infty} \frac{1}{n^3} = \zeta(3)\,|\psi_{1S}(0)|^2,
\end{equation}
where $\zeta(3)\approx 1.202$ is Ap\'ery's constant, yielding a total rate $\Gamma_{\text{total}} \approx 1.2 \times \Gamma(1S)$.

\subsection{Yield estimates at STCF, RHIC, and the LHC}

Once the branching ratio for $D^0\to (K\mu)_{\rm atom}\nu_\mu$ is obtained, the expected atom yield at a given facility follows from the produced $D^0$ yield, multiplied by the corresponding branching ratio and by the relevant acceptance and efficiency factors. 
For hadronic collisions, the $D^0$ abundance is estimated from the charm-production cross section, the fragmentation fraction $f(c\to D^0)$, and, for nucleus--nucleus collisions, binary-collision scaling with the average number of nucleon--nucleon collisions $\langle N_{\rm coll}\rangle$. 
For $e^+e^-$ collisions at STCF, the estimate is based on the expected integrated luminosity and the projected $D^0\bar D^0$ production rate. 
These inputs are then used in Sec.~III to compare the production prospects in RHIC Au+Au, LHC p+p and Pb+Pb, and STCF running scenarios.

\subsection{Dissociation in detector material and observable signatures}

A characteristic feature of a neutral muonic kaon atom is its dissociation in detector material. Once formed, the $(K\mu)_{\rm atom}$ can be ionized or broken up through interactions with nuclei or electrons in the beampipe and inner tracking layers. The dissociation probability is estimated using the standard exponential attenuation formula
\begin{equation}
f = 1 - e^{-n\sigma l},
\end{equation}
where $f$ is the dissociation fraction, $n$ is the number density of the material, $\sigma$ is the atom--material interaction cross section, and $l$ is the material thickness. 
Following Ref.~\cite{Mrowczynski:1985qt}, we use dissociation cross sections of those for $\pi K$ atoms, which are close in size to the $K\mu$ system: $\sigma=140$ barn with Carbon as target and $\sigma=650$ barn with Al. 
Detector material budgets—including beampipes, support structures, and inner tracking layers—are well documented for STAR, CMS, and other experiments. These material properties (density, composition, thickness) allow for detailed calculations of dissociation probability~\cite{Matis:2002se,Anderson:2003ur,GoyLopez:2017kkg,CMS:2012sda}.
The resulting dissociation probabilities and their implications for secondary-vertex reconstruction are discussed in Sec.~III~E.

\section{Results and Discussion}

\subsection{Branching ratios for atomic channels}

Analogously to the free case, the branching ratios for the atomic channels are
\begin{align}
\label{atomBR}
\text{BR}\big(D^0 \to (K\mu)_{1S}\,\nu_\mu\big)
&=
\Gamma\big(D^0 \to (K\mu)_{1S}\,\nu_\mu\big)\,
\frac{\tau_{D^0}}{\hbar},
\\
\text{BR}\big(D^0 \to (K\mu)_{nS\,\text{all}}\,\nu_\mu\big)
&\approx
1.2\;\text{BR}\big(D^0 \to (K\mu)_{1S}\,\nu_\mu\big).
\end{align}
Numerically, the branching ratios for the $1S$ state and for the sum over all $nS$ states are found to be
\[
\begin{aligned}
\text{BR}\big(D^0 \to (K\mu)_{1S}\,\nu_\mu\big) &= 1.91\times 10^{-10},\\
\text{BR}\big(D^0 \to (K\mu)_{nS\,\text{all}}\,\nu_\mu\big) &= 2.29\times 10^{-10}.
\end{aligned}
\]
These values are tiny compared with those of the free three-body decay $D^0\to K^-\mu^+\nu_\mu$, but they provide the quantitative basis for assessing whether the process can nevertheless be accessed at modern high-luminosity collider facilities.

\subsection{Production yields at  RHIC, the LHC, and STCF}

Using Eq.~\eqref{atomBR}, the expected yields of muonic kaon atoms can be estimated for RHIC, the LHC, and STCF~\cite{STCF:2025xop,Achasov:2023gey}. The results are summarized in Table~\ref{table:yields}.

For RHIC, the STAR experiment collected a total of about 20 billion Au+Au collision events at $\sqrt{s_{\rm NN}}=200$ GeV in the 2023 and 2025 runs~\cite{starBUR}. 
The charm-pair production cross section in p+p collisions at 200 GeV has been measured to be $\sigma_{c\bar c}=170~\mu{\rm b}$~\cite{STAR:2012nbd}. Combined with the fragmentation fraction $f(c\to D^0)=0.565$~\cite{STAR:2012nbd}, this gives a $D^0$ production cross section of
\[
\sigma_{D^0}=\sigma_{c\bar c}\times f(c\to D^0)=170~\mu{\rm b}\times 0.565 = 96.0~\mu{\rm b}.
\]
With the proton--proton inelastic cross section $\sigma_{pp}^{\rm inel}=42~{\rm mb}$~\cite{Loizides:2017ack} and the average number of binary collisions for the 0--100\% centrality range, $\langle N_{\rm coll}\rangle=235.1$~\cite{Loizides:2017ack}, the total $D^0$ yield in Au+Au collisions is estimated as
\[
N_{D^0} = \frac{\sigma_{D^0}}{\sigma_{pp}^{\rm inel}} \times \langle N_{\rm coll}\rangle \times N_{\rm evt}
\approx 1.08\times 10^{10}.
\]
This translates into only a few $(K\mu)_{\rm atom}$ candidates from the decay channel. Specifically, the expected yield of $K^+\mu^-$ or $K^-\mu^+$ atomic pairs from $D^0$ decays is about 2.5, which is far smaller than the production expected from QGP coalescence~\cite{Wang:2024njn}.

At the LHC, high-luminosity p+p running provides by far the most favorable environment for decay-driven production. Taking the projected CMS integrated luminosity of 3000 fb$^{-1}$ by 2041 as a benchmark, and using the $c\bar c$ production cross section of 1347.4~$\mu$b measured at $\sqrt{s}=7$ TeV~\cite{ALICE:2021dhb} as the baseline input, one obtains a total yield of about $5.23\times 10^5$ atoms, together with an equal number of anti-atoms. Even under a conservative assumption of an overall detection efficiency of only $1\%$, one would still expect on the order of $5\times10^3$ detected atoms.

For Pb+Pb collisions at the LHC, the integrated luminosity collected by CMS during Run~2 and Run~3 reaches $8550~\mu{\rm b}^{-1}$~\cite{CMSluminosity}. Using $\langle N_{\rm coll}\rangle=393$ for 0--100\% Pb+Pb collisions at $\sqrt{s_{\rm NN}}=5.02$ TeV~\cite{CMS:2016xef}, the expected yield from the $D^0$ decay channel is only of order unity. 
For STCF, the conceptual design report projects an integrated luminosity of about 1~ab$^{-1}$ over 20 years of operation~\cite{Achasov:2023gey}. Using the anticipated $D^0\bar D^0$ production cross section, the total number of produced $(K\mu)_{\rm atom}$ states is estimated to be $\mathcal{O}(10)$, which is challenging but still potentially interesting because of the exceptionally clean $e^+e^-$ environment.

\begin{table*}[t]
\caption{Total muonic atom yields at different facilities. RHIC Au+Au collisions correspond to STAR/sPHENIX running. LHC p+p assumes an effective trigger strategy for charm production. STCF assumes the integrated luminosity accumulated over 20 years of operation. The QGP-coalescence yields are taken from Ref.~\cite{Wang:2024njn}.}
\begin{tabular*}{\textwidth}{@{\extracolsep{\fill}}|c|c|c|c|c|@{}}
\hline
  & RHIC (Au+Au, 200 GeV) & LHC (p+p, 7 TeV) & LHC (Pb+Pb, 2.76-5.36 TeV) & STCF\\ 
\hline
Luminosity & 2000--5000 $\mu b^{-1}$  & 3000 $fb^{-1}$ & 8550 $\mu b^{-1}$ & 1 $ab^{-1}$\\
\hline
$D^0$ counts & $10^{10}$ & $2 \times 10^{15}$ & $1 \times 10^{12}$ & $8 \times 10^{10}$\\
\hline
$(K^{\pm}\mu^{\mp})_{\rm atom}$ from $D^0$ decay & $\sim10$ & $5 \times 10^{5}$ & $\sim200$ & $\sim20$\\
\hline
$(K^{\pm}\mu^{\mp})_{\rm atom}$ from QGP coalescence & $10^5$ & N/A & $10^5$ & N/A\\
\hline
\end{tabular*}
\label{table:yields}
\end{table*}

\subsection{Comparison with coalescence production in heavy-ion collisions}

The branching ratio for the decay channel
\[
D^0 \to (K\mu)\,\nu_\mu
\]
is of order $10^{-10}$, which is roughly seven orders of magnitude smaller than that of the free three-body decay
\[
D^0 \to K^- \mu^+ \nu_\mu.
\]
This strong suppression is expected, since the kaon and muon must be produced with extremely small relative momentum in order to bind into a Coulombic state. 
Nevertheless, the very high luminosities available at modern collider facilities still make the process experimentally relevant, especially in LHC p+p running.

The decay-driven mechanism should be contrasted with coalescence production in the bulk medium of heavy-ion collisions. In that case, the yield of muonic atoms can be written as~\cite{Wang:2024njn}
\begin{equation}
\frac{dN_{\rm atom}}{dyd^2p_{T,atom}} = \Big(\frac{4\pi}{3}(\delta p_{_{\rm QGP}})^3\Big)\frac{1}{m_{\rm red}}\frac{dN_h}{dyd^2p_{T,h}}\frac{dN_l}{dyd^2p_{T,l}},
\label{eq:coal}
\end{equation}
where ${dN_x}/{dyd^2p_{T,x}}$ denotes the spectra of the hadron and lepton constituents, $\delta p_{\rm QGP}$ is the relative-momentum cutoff for coalescence, $m_{red} = \frac{m_{h}m_{l}}{m_{h}+m_{l}}$ is the reduced mass, $m_{h}$ and $m_{l}$ denote the mass of the hadron and lepton constituents.

The coalescence cutoff is estimated as~\cite{Wang:2024njn}
\begin{equation}
\delta p_{_{\rm QGP}} \leq \sqrt{2m_{\rm red}\frac{1}{4\pi\varepsilon_0}\frac{e^2}{r_0}} \simeq 5\text{--}10~\text{MeV}/c,
\label{space}
\end{equation}
where $r_0\simeq 3$--$10$ fm is the correlation length in the approximately homogeneous freeze-out volume of A+A collisions~\cite{Wang:2024njn,Sun:2018jhg,Hillmann:2021zgj,Butler:1961pr,Nagle:1996vp,Llope:1995zz}.

For STAR at RHIC, the detector momentum resolution around a kaon momentum of 1~GeV/$c$ is approximately $\delta p_{\rm det}\simeq 1.5-3\%\,p \simeq 15-30$ MeV/$c$~\cite{Anderson:2003ur}. This implies a characteristic signal-to-background ratio
\begin{equation}
\Big(\frac{\delta p_{_{\rm QGP}}}{\delta p_{\rm det}}\Big)^3 \simeq 1/200\text{--}1/10.
\end{equation}
In practice, Coulomb correlations near threshold further increase the background by about a factor of 2~\cite{Xin:2014lza,STAR:2003cqe}, making the search challenging, though still comparable in difficulty to thermal dielectron measurements~\cite{STAR:2024bpc}.

The physics information carried by the two mechanisms is complementary. 
Decay-driven production probes weak-decay dynamics and the near-threshold projection of a channel-defined $K\mu$ pair, whereas coalescence production probes thermal momentum distributions, freeze-out correlations, and the early-time electromagnetic emissivity of the QGP. 
In heavy-ion collisions, the coalescence contribution clearly dominates the total $(K\mu)_{\rm atom}$ yield, while in p+p collisions the decay channel becomes the relevant production source.

\subsection{$K\mu$ atoms as a probe of primordial low-$p_T$ muons}

Because the quark--gluon plasma exists for only a few fm/$c$, it cannot be observed directly. Its properties must therefore be inferred from final-state particles recorded by detectors. Stable hadrons are formed only during the late hadronic stage, which limits their sensitivity to the earliest QGP phase~\cite{STAR:2005gfr}. By contrast, leptons are produced throughout the full space--time evolution of the collision and escape the medium with minimal final-state interactions, making them especially valuable electromagnetic probes.

\begin{figure}
\includegraphics[width=0.5\textwidth]{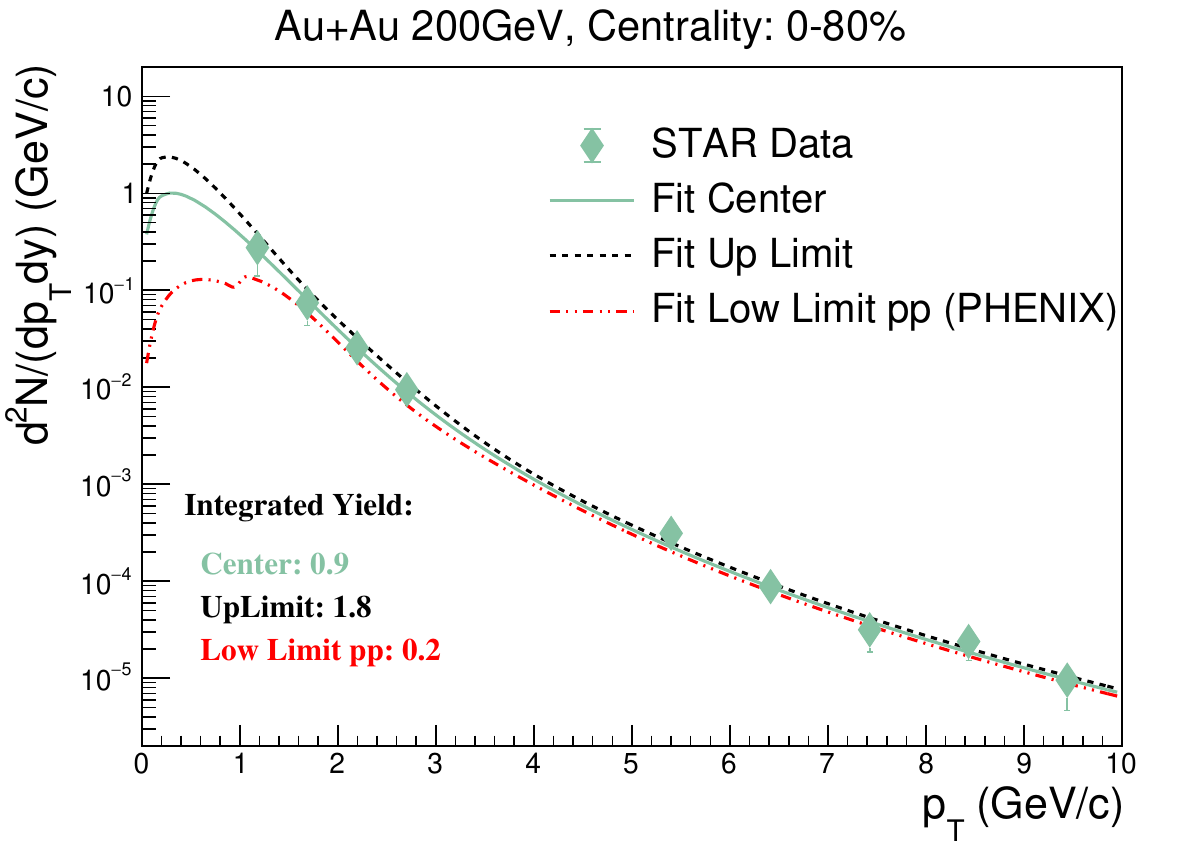}
\caption{Exponential-function fits to the STAR direct virtual-photon $p_T$ spectrum~\cite{STAR:2016use} are used to extrapolate into the low-$p_T$ region where no direct experimental measurements exist. The integrated yield and its upper/lower limits are indicated in the figure in the same color as the fit line.}
\label{fig:mu_extrap}
\end{figure}

Experimental access to the QGP temperature remains limited. Dilepton and direct virtual-photon measurements are among the few observables capable of constraining thermal radiation from the plasma, and major experiments including NA60, STAR, PHENIX, and ALICE have pursued such studies~\cite{STAR:2024bpc,STAR:2016use,PHENIX:2022rsx}. 
However, in the low-transverse-momentum region of the direct virtual-photon spectrum, where thermal radiation is strongest, the absence of direct experimental constraints requires the measured high-$p_T$ data to be fitted and extrapolated. Following Ref.~\cite{STAR:2016use}, we use
\begin{equation}
Ae^{-p_{T}/T} + \frac{N_{coll}^{STAR}}{\sigma_{pp}} \times A_{pp}(1+p_{T}^{2}/b)^{-n}
\label{eq:directPhotonFitFunction}
\end{equation}
to describe the direct virtual-photon spectrum. Figure~\ref{fig:mu_extrap} shows a schematic extrapolation of the STAR direct virtual-photon $p_T$ spectrum into the low-$p_T$ region. The figure highlights the central issue relevant for the present study: while the measured spectrum at moderate and high $p_T$ is reasonably constrained, the low-$p_T$ region---which is most sensitive to thermal radiation from the QGP---still suffers from substantial uncertainty. Below $p_T =$ 1 GeV/$c$, the lower limit from the fit drops sharply to zero. Therefore, for $p_T <$ 1 GeV/$c$, the lower limit shown in the figure is taken from the $N_{\rm coll}$-scaled p+p result~\cite{PHENIX:2022rsx,PHENIX:2009gyd}, which results in an apparent discontinuity around 1 GeV/$c$. Even when the $N_{\rm coll}$-scaled p+p result is used as a conservative lower bound~\cite{PHENIX:2022rsx,PHENIX:2009gyd}, the extrapolated low-$p_T$ yield can still differ from the central estimate by roughly an order of magnitude. This large uncertainty motivates the search for complementary observables that are directly sensitive to primordial low-$p_T$ muons.

\begin{figure}[t]
    \centering
    \includegraphics[width=0.48\textwidth]{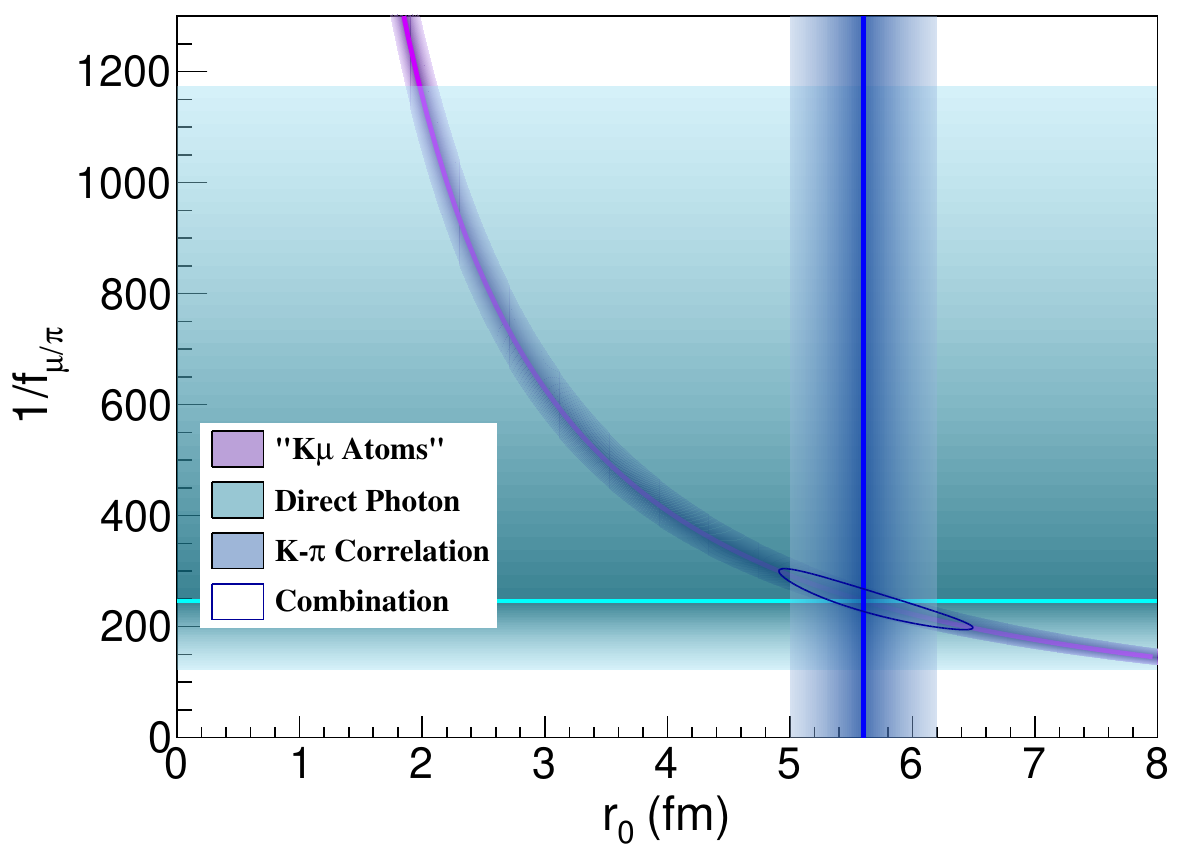}
    \caption{
        Schematic dependence of the $\pi$--$\mu$ yield ratio $1/f_{\mu/\pi}$ as a function of the coalescence radius $r_0$. 
        The cyan band shows the value of $1/f_{\mu/\pi}$ and its $1\sigma$ interval derived from the direct virtual-photon spectrum in Fig.~\ref{fig:mu_extrap}, where extrapolation into the low-$p_T$ region introduces significant uncertainty. 
        The blue band represents the constraint on $r_0$ from the $K$--$\pi$ correlation analysis~\cite{STAR:2003cqe}. 
        The purple band indicates a projected yield and $1\sigma$ interval obtained by combining an assumed $(K\mu)_{\rm atom}$ yield with Eq.~\eqref{eq:coal} and an assumed $\pm10\%$ uncertainty. 
        The overlap of the three bands provides a strong combined constraint on the primordial muon yield.
    }
    \label{fig:f2r0}
\end{figure}

Muonic kaon atoms provide such an alternative handle. According to Eq.~\eqref{eq:coal}, the coalescence yield of $(K\mu)_{\rm atom}$ is proportional to the yields of the constituent hadrons and primordial muons. A measurement of the atom yield therefore provides direct access to the primordial low-$p_T$ muon abundance and can thus tighten constraints on the theoretical description of thermal electromagnetic radiation.
Figure~\ref{fig:f2r0} schematically illustrates the constraints of the pion-to-muon yield ratio $1/f_{\mu/\pi}$ and the coalescence radius $r_0$. 
The cyan band corresponds to the value of $1/f_{\mu/\pi}$ and its $1\sigma$ interval derived from the direct virtual-photon spectrum shown in Fig.~\ref{fig:mu_extrap}. To obtain the primordial muon spectrum, one can perform a Monte Carlo simulation of direct-photon decays using the transverse-momentum spectrum in Fig.~\ref{fig:mu_extrap} together with the invariant-mass distribution in Ref.~\cite{STAR:2016use}. The pion yield can be taken from Tsallis--Blast-Wave fits to Au+Au collisions at \sNN $=$ 200 GeV~\cite{Chen:2020zuw}.  
The asymmetric uncertainty originates solely from the muon yield uncertainty shown in Fig.~\ref{fig:mu_extrap}. Although the asymmetry in the muon yield uncertainty is very small, the muon yield appears in the denominator on the y‑axis of Fig.~\ref{fig:f2r0}, which amplifies the apparent asymmetry of the uncertainty. The blue band represents the constraint on $r_0$ from the $K$--$\pi$ correlation analysis~\cite{STAR:2003cqe}. 
The purple band shows a projected constraint obtained from a hypothetical $(K\mu)_{\rm atom}$ yield measurement with a representative $\pm10\%$ uncertainty. The overlap of the cyan, blue, and purple bands in Fig.~\ref{fig:f2r0} then defines a combined constraint on the primordial muon yield. Specifically, using the value of the cyan solid line, the value of the blue solid line, and Eq. (22), we derive a hypothetical central value for the muonic atom yield, which is depicted as the purple solid line in the figure. A 10\% uncertainty is subsequently assigned to this central value, represented by the purple band.
This makes $(K\mu)_{\rm atom}$ measurements a potentially powerful tool for constraining the low-$p_T$ region of thermal dilepton and direct-photon production, a region that remains terra incognita for present experiments.

\subsection{Experimental feasibility}

A characteristic signature of a muonic kaon atom is its dissociation while traversing detector material. Once formed, the neutral atom can be broken up through interactions with nuclei or electrons in the beampipe and inner tracking layers, producing a correlated $K$--$\mu$ pair with a very small relative velocity. 
Using the attenuation formula introduced in Sec.~II~D and the dissociation cross sections from Ref.~\cite{Mrowczynski:1985qt}, together with the material budgets of STAR and CMS~\cite{Matis:2002se,Anderson:2003ur,GoyLopez:2017kkg,CMS:2012sda}, one obtains sizable breakup probabilities already in the first layers of the detector. These probabilities are detailed in Table~\ref{table: dissociation fraction}.

For STAR, the beryllium beampipe~\cite{Matis:2002se} ($Z=4$, $n = 1.236\times10^{23}$ atoms/cm$^3$) with a thickness of 0.1 cm yields a dissociation probability of about $53.5\%$. 
For CMS, the slightly thinner beryllium beampipe (0.08 cm) gives a dissociation probability of about $45.8\%$. 
After traversing the beampipe, the surviving atoms pass through air gaps and the first-layer tracking material, where nearly all remaining atoms dissociate. 
In practice, this means that the total breakup fraction is close to $100\%$ before the atom reaches the deeper tracking systems.

\begin{table*}[htp]
\caption{Estimated dissociation fractions of $(K\mu)_{\rm atom}$ in representative detector materials for STAR and CMS.}
\begin{center}
\begin{tabular}{|c|c|c|}
\hline
  & RHIC-STAR & LHC-CMS \\ 
\hline
Beampipe $(Z,n,l)$ & Be$(4,\,1.236 \times 10^{23}\,{\rm atoms/cm^3},\,0.1~{\rm cm})$ & Be$(4,\,1.236 \times 10^{23}\,{\rm atoms/cm^3},\,0.08~{\rm cm})$ \\
\hline
Dissociation at beampipe & $53.5\%$ & $45.8\%$ \\
\hline
Air $(Z,n,l)$  & $(7.2,\,2.54 \times 10^{19}\,{\rm atoms/cm^3},\,46.19~{\rm cm})$ & *** \\
\hline
Dissociation in air & $(1-53.5\%)\times 20.9\%$ & *** \\
\hline
First-layer detector $(Z,n,l)$ & Inner field cage (Kapton, $0.6\% X_0$) & Si$(14,\,5.0 \times 10^{22}\,{\rm atoms/cm^3},\,13.1~{\rm cm})$ \\
\hline
Total dissociation before deeper tracking & $\simeq 100\%$ & $\simeq 100\%$ \\
\hline
\end{tabular}
\end{center}
\label{table: dissociation fraction}
\end{table*}

This large dissociation probability is experimentally advantageous because the breakup products originate from a secondary vertex downstream of the primary $D^0$ decay point. The emerging kaon and muon are nearly comoving and therefore exhibit a small opening angle and a distinctive low invariant mass. These topological and kinematic features provide the main experimental handle for identifying the signal.

A simple phase-space estimate illustrates why vertex separation is essential. Following Eq.~(1) and Eq.~(3) of Ref.~\cite{Wang:2024njn}, the atom yield can be written as
\begin{equation}
\begin{aligned} 
\frac{dN_{\rm atom}}{dyd^2p_{T,{\rm atom}}}
= 8\pi^2\zeta (3) \alpha^3 m^2_{\rm red}
\frac{dN_h}{dyd^2p_{T,h}} \frac{dN_l}{dyd^2p_{T,l}}\\
=\Big( \frac{4\pi}{3}(\delta p_{a})^{3} \Big) \, \frac{1}{m_{\rm red}}
\frac{dN_h}{dyd^2p_{T,h}} \frac{dN_l}{dyd^2p_{T,l}} ,
\label{eq:dpa}
\end{aligned}
\end{equation}
which corresponds to an effective coalescence momentum of $\delta p_a \simeq 1.8$ MeV/$c$. 
For unbound kaons from $D^0$ decay, the relevant momentum range is $\delta p_{\rm dec}\lesssim 800$ MeV/$c$, so the approximate phase-space suppression factor is
\begin{equation}
\frac{\Gamma(D^0\to(K\mu)_{a}\nu)}{\Gamma(D^0\to K\mu\nu)}
\sim \Big(\frac{\delta p_a}{\delta p_{\rm dec}}\Big)^3 \simeq 10^{-8}.
\end{equation}
This estimate is consistent with the more rigorous branching-ratio calculation based on Eqs.~\eqref{eq:Gamma_atom_final} and \eqref{atomBR}.

Without vertex separation between the secondary vertex from $D^0$ decay and the subsequent tertiary vertex from atom dissociation at beampipe and detector layer, the expected signal-to-background ratio would be only
\begin{equation}
\Big(\frac{\delta p_a}{\delta p_{\rm det}}\Big)^3 \simeq 10^{-3},
\end{equation}
which is clearly too small for a practical search. 
By contrast, a displaced breakup tertiary vertex, together with the characteristic near-comoving $K$--$\mu$ topology, provides a realistic path toward background suppression. 
With modern inner trackers such as ALICE ITS2/ITS3 and the CMS Phase-2 tracker, and with the large luminosities expected in future p+p and A+A running, the observation of $(K\mu)_{\rm atom}$ states appears feasible, especially for the high-statistics LHC p+p program and, for coalescence production, in heavy-ion collisions at RHIC and the LHC.

\section{Summary}

We have developed a quantitative theoretical framework for the production of muonic kaon atoms in the semileptonic decay channel 
$D^0 \to (K\mu)_{\rm atom}\,\nu_\mu,$
combining the weak decay amplitude, the three-body phase-space treatment, and the nonrelativistic bound-state projection into a single consistent description. 
This framework yields branching ratios of order $10^{-10}$ for the atomic channel, with
\[
\begin{aligned}
\text{BR}\big(D^0 \to (K\mu)_{1S}\,\nu_\mu\big) &= 1.91\times 10^{-10},\\
\text{BR}\big(D^0 \to (K\mu)_{nS\,\text{all}}\,\nu_\mu\big) &= 2.29\times 10^{-10}.
\end{aligned}
\]

Using these branching ratios, we estimated the expected yields at RHIC, the LHC, and STCF. 
The decay-driven contribution is negligible in heavy-ion collisions compared with the much larger coalescence production in the QGP, whereas high-luminosity p+p running at the LHC provides a particularly favorable environment for observing atoms produced in $D^0$ decays. 
At the same time, the coalescence channel in A+A collisions remains especially interesting because of its sensitivity to the primordial low-$p_T$ muon yield and thus to the thermal electromagnetic emissivity of the QGP.

We further incorporated atom dissociation in detector material into the analysis. 
The estimated breakup probabilities in the beampipe and first tracking layers are large, implying that most produced atoms will dissociate into nearly comoving correlated $K$--$\mu$ pairs before reaching the outer detector. 
This leads to a distinctive experimental signature characterized by displaced breakup vertices, small opening angles, and low invariant masses.
Overall, our results show that muonic kaon atoms are not only theoretically well defined in their formation but also experimentally accessible at future high‑statistics facilities.
They thus represent a promising new probe of exotic bound‑state formation, weak charm decays, and the electromagnetic radiation dynamics of the QGP.

\section{Acknowledgments}
This work was supported in part by the National Natural Science Foundation of China under Grant Nos. 12361141827 and 12422510, and by the Office of Nuclear Physics within the U.S. Department of Energy Office of Science under Contract DE‑FG02‑89ER40531. The National Key Research and Development Program of China provided additional support under Contract No. 2022YFA1604900.
W. Zha acknowledges support from the Anhui Provincial Natural Science Foundation (Grant No. 2508085JX002), the Youth Innovation Promotion Association of the Chinese Academy of Sciences, and the Chinese Academy of Sciences under Grant No. YSBR088.

\bibliography{apssamp}

\end{document}